\documentclass{svjour3}                     
\smartqed  
\usepackage{graphicx}
\newcommand{\vf}{v_{\rm F}}

\newcommand{\ef}{\epsilon_{\rm F}}

\newcommand{\kf}{k_{\rm F}}

\begin{document}

\title{Models of electron transport in single layer graphene
}


\author{F. Guinea
}


\institute{F. Guinea \at
              Instituto de Ciencia de Materiales de Madrid. Sor  Juana In\'es de la Cruz 3. E-28049 Madrid. Spain
              and  Donostia International Physics Center (DIPC), Paseo Manuel de
Lendiz\'abal 4, San Sebasti\'an, E-20018, Spain \\
              \email{paco.guinea@icmm.csic.es}           
}

\date{Received: date / Accepted: date}

\maketitle

\begin{abstract}
The main features of the conductivity of doped single layer graphene
are analyzed, and models for different scattering mechanisms are
presented. Many possible dependencies of the cross section on the
Fermi wavelength are identified, depending on the type of scattering
mechanism. Defects with internal structure, such as ripples, show
non monotonous dependencies, with maxima when the Fermi wavelength
is comparable to the typical scale of the defect.
\end{abstract}
\section{Introduction.}
A great research effort devoted to graphene started after the
realization that single layer graphene can be isolated and that the
number of carriers can be tuned\cite{Netal04,Netal05}. The
properties of epitaxially grown samples with few graphene
layers\cite{Betal04}, has also induced a significant activity. Many
of the potential applications are related to the design of
electronic devices.

We analyze here some topics related to the electronic transport
properties of single layer graphene. We do not consider models for
the carrier mobility in systems with more than one layer. It is
interesting to note in this respect that the mobility of graphite is
significantly higher than that of few layer graphene
samples\cite{GEBMS08}. We will also not analyze the interesting
topic of the transport properties of undoped graphene.

We restrict ourselves here to doped single layer samples in the
diffusive regime, where the mean free path is shorter than the
sample size. In this regime, the motion of the carriers can be
described semiclassically, and the carrier mobility is determined by
the different scattering mechanisms present in the system. If we
describe the effects of the scattering events on the carrier
distribution function using the Boltzmann equation, which is a
reasonable approximation in this limit, the conductivity of graphene
is given by:
\begin{equation}
\sigma = \frac{e^2}{h} \vf^2 N ( \ef ) \tau ( \ef )
\label{conductivity}
\end{equation}
where $N ( \ef )$ is the density of states, and $\tau ( \ef )$ is
the scattering time.

We present, in the following section, a brief description of the
behavior of the conductivity in single layer graphene, as function
of carrier concentration and temperature. Then, we analyze different
scattering mechanisms which may limit the mobility at low
temperatures.
\section{Qualitative behavior of the conductivity in graphene.}
\subsection{Low temperature limit.}
It was realized at an early stage that the conductivity at low
temperatures was almost directly proportional to the carrier
density\cite{Netal04}, and this dependence has been repeatedly
confirmed\cite{Tetal07}. As, $N ( \ef ) \propto \sqrt{\rho}$ in
graphene, where $\rho$ is the carrier density, this behavior implies
that $\tau ( \ef ) \propto \sqrt{\rho}$. On the other had, for weak
local scatterers the Born approximation predicts an inverse
scattering time proportional to the density of states, $\tau^{-1}
\propto n_{imp} N ( \ef ) \propto \sqrt{\rho}$, where $n_{imp}$ is
the number of scatterers. Inserting this expression in
eq.(\ref{conductivity}), we find that weak scatterers lead to a
conductivity which is independent of the carrier density, in
contradiction with the observed behavior.

The linear dependence on carrier density of the conductivity implies
that the scattering time should increase with density as $\tau (
\rho ) \propto \sqrt{\rho}$, or, analogously, the cross section of
the defects, which in two dimensions is given by a length, should
scale as $\kf^{-1}$.

The first mechanism proposed compatible with a $\tau \propto
\sqrt{\rho}$ dependence was scattering by charged
impurities\cite{NM07,AHGS07}. The Coulomb potential is scale
invariant, as it only depends on the product of the charge of the
impurity and the electron charge, $Z e^2$, which can be rendered
dimensionless by dividing it by the Fermi velocity. Hence, the cross
section should be proportional to the only length scale in the
problem, which is the Fermi wavelength, $\lambda_{\rm F} =
\kf^{-1}$. This dimensional argument\cite{N08} remains valid even
when screening by the carriers is included, as the Fermi-Thomas
screening length is proportional to the Fermi wavelength in
graphene. The electronic structure of graphene near a charged
impurity has been studied extensively in later
times\cite{SKL07,FNS07,PNN07}, without changing the previous
analysis.

A different mechanism which leads to a linear dependence on the
Fermi wavelength of the scattering cross section is induced by
strong scatterers, such as lattice vacancies\cite{PGLPN06}. These
defects change significantly the local density of states, as they
induce partially localized states at the Dirac energy. As reviewed
below, the scattering phasehifts cannot be described by the Born
approximation\cite{KN07,HG07,SPG07}.

Other mechanisms which lead to deviations from the $\tau^{-1} ( \ef
) \propto N ( \ef )$ dependence expected from the Born approximation
are scattering by ripples\cite{KG08} (see also below), and
scattering by defects which lead to long range distortions of the
lattice, such as dislocations.
\subsection{Finite temperature conductivity}
The resistivity of single layer graphene rises as function of
temperature\cite{TZSK07,Metal08,CJXIF08}, and deviates significantly
from its low temperature value at room temperature.

The coupling to in plane phonons is well understood\cite{G81,HS07}.
The phonon band width is $\sim 0.2$eV. The number of thermally
excited phonons at room temperature is too small to explain the
observed rise in the resistivity. Single layer graphene can also
support out of plane flexural modes, which show a high density of
states at low energies. The coupling to the electrons is quadratic
on the phonon coordinates, and the resulting scattering rate is also
too low to explain the observed temperature dependence of the
resistivity\cite{MO08}. Note that these modes can be pinned by the
substrate\cite{SSFGNS08}, reducing their density of states at low
energies.

The most likely explanation for the temperature dependence of the
resistivity is scattering by substrate modes\cite{FG08}. The most
common substrate used in experiments on graphene samples obtained by
mechanical cleavage is SiO$_2$, which is a polar insulator. The
electrons in graphene couple to the electric fields induced by the
surface polar modes. These modes can be thermally excited at room
temperature. A fit using the observed frequencies and dielectric
function of SiO$_2$ gives a good agreement with experimental
data\cite{CJXIF08}.
\subsection{Effects of the substrate}
Localized charges in the substrate lead to Coulomb scattering and
modify the low temperature mobility. At finite temperatures, the
conductivity depends on the coupling of the carriers to the
substrate modes. Hence, the transport properties of single layer
graphene are determined, to a large extent, by the substrate and by
the general properties of the surrounding environment. Recent
experiments show that the conductivity in graphene is modified in
samples suspended above the substrate\cite{Betal08,DSBA08}. The
presence of water molecules can screen charged
impurities\cite{Setal07b}, as well as change the adhesion between
the graphene layer and the substrate\cite{MVJBB08,SSFGNS08}. A
detailed investigation of the effects of different substrates will
be very helpful for the understanding of the carrier transport in
graphene.
\section{Scattering processes in graphene}
\subsection{General framework}
In the following, we analyze different scattering mechanisms which
may be present in graphene. We first discuss the generalization of
the standard partial wave analysis of scattering off local
potentials in quantum mechanics to Dirac quasiparticles\cite{LL58}.
The formalism is rather general, and has already been formulated,
using different notation, in\cite{N07}. Related results can be found
in\cite{KN07,HG07}. We then apply it to different scattering
processes. The continuum model used neglects scattering between the
two inequivalent valleys in graphene. We finally give a scheme which
takes into account explicitly the lattice structure, and which can
be used to study scattering processes where intervalley scattering
is significant.

We analyze the phaseshifts induced by a circular potential well in
graphene. Using cylindrical coordinates, the Hamiltonian in the
clean system can be written as:
\begin{equation}
{\cal H} \equiv \vf \left( \begin{array}{cccc} 0 &i e^{- i \phi}
\partial_r +
    \frac{e^{- i \phi}}{r} \partial_\phi &0 &0 \\ i e^{i \phi} \partial_r -
    \frac{e^{- i \phi}}{r} \partial_\phi &0 &0 &0 \\ 0 &0 &0 &- i e^{i \phi}
    \partial_r + \frac{e^{i \phi}}{r} \partial_\phi \\ 0 &0 &- i e^{-i \phi}
    \partial_r - \frac{e^{-i \phi}}{r} \partial_\phi &0
\end{array} \right)
\label{hamil}
\end{equation}
where the two first entries correspond to the $K$ point, and the two
last ones to the $K'$ point, and we are setting the Fermi velocity
$\vf=1$.

In the following, we study scattering processes which do not induce
intervalley transitions. Hence, we need only consider one valley. We
write the incoming wave as:
\begin{equation}
\Psi_{in} ( {\bf \vec{r}} ) \equiv \left( \begin{array}{c} 1 \\ e^{i
      \theta_k} \end{array} \right) e^{i {\bf \vec{k}} {\bf
      \vec{r}}}  \equiv \left( \begin{array}{c} \sum_n i^n J_n ( k r )
      e^{i n \theta} e^{-i  n \theta_k } \\ \sum_n i^{n+1} J_{n+1} ( kr )
      e^{i ( n+1 )
      \theta}  e^{-i  n \theta_k }  \end{array} \right)
\label{wv_in}
\end{equation}
where the angle $\theta_k$ defines the direction of the vector ${\bf
  \vec{k}}$. In the following, we will set $\theta_k=0$. The incoming current
  is $j_x^{in} = k , j_y^{in} = 0$.

The outgoing wave is:
\begin{eqnarray}
\lim_{| \vec{r} | \rightarrow \infty} \Psi_{out} ( {\bf \vec{r}} )
&= & \left(
    \begin{array}{c} f ( \theta ) \\ f
    ( \theta ) e^{i \theta} \end{array} \right) \sqrt{\frac{2}{\pi k r}} e^{i
    k r} \approx \nonumber \\ &\approx &
    e^{i \pi / 4} \left(
    \begin{array}{c} \sum_n i^n f_n \left[ J_n ( k r ) + i Y_n ( k r ) \right]
     e^{i n \theta} \\  \sum_n i^{n+1} \left[ f_n [ J_{n+1} ( kr )
    + i Y_{n+1} ( k r
    ) \right] e^{i ( n+1 ) \theta}  \end{array} \right)
\label{wv_out}
\end{eqnarray}
The outgoing current is $j_x^{out} = \langle \Psi_{out} | \sigma_x |
\Psi_{out} \rangle = 2 | f ( \theta ) |^2 \cos ( \theta ) ,
j_y^{out} = \langle \Psi_{out} | \sigma_y | \Psi_{out} \rangle = 2 |
f ( \theta )  |^2 \sin ( \theta )$. The differential cross section
is $d \sigma (  \theta ) = ( 4 | f ( \theta ) |^2 ) / ( \pi k ) d
\theta$.

In order to obtain eq.(\ref{wv_out}),  we have made the ansatz:
\begin{equation}
f ( \theta ) = \sum_n f_n e^{i n \theta} \label{angular_amplitude}
\end{equation}
and we have used the expansion:
\begin{eqnarray}
\lim_{kr \rightarrow \infty} J_n ( k r ) &\approx
&\sqrt{\frac{2}{\pi k r}}
    \cos \left( k r - \frac{n
    \pi}{2} - \frac{\pi}{4} \right) \nonumber \\
\lim_{kr \rightarrow \infty} Y_n ( k r ) &\approx
&\sqrt{\frac{2}{\pi k r}}
    \sin \left( k r - \frac{n
    \pi}{2} - \frac{\pi}{4} \right)
\end{eqnarray}

We assume that the scattering defect has a finite radius, $r_0$. We
can write the wavefunctions of an electron with energy $\vf k$
outside this radius as a superposition of terms:
\begin{equation}
\Psi_n ( {\bf \vec{r}} ) \equiv \left( \begin{array}{c} \left[ J_n (
k r ) +
      R_n Y_n ( k r ) \right] e^{i n \theta} \\ i \left[ J_{n+1} + R_n
      Y_{n+1} (
      k r ) \right] e^{i ( n+1) \theta} \end{array} \right)
\label{solution}
\end{equation}
We now write:
\begin{equation}
\Psi_{in} ( {\bf \vec{r}} ) + \Psi_{out} ( {\bf \vec{r}} ) = \sum_n
\alpha_n \Psi_n ( {\bf \vec{r}} )
\end{equation}
And, using  eq.(\ref{wv_in}), eq.(\ref{wv_out}) and
eq.(\ref{solution}), we obtain:
\begin{eqnarray}
f_n &= &\frac{R_n e^{-i \pi / 4}}{i - R_n} \nonumber \\
f ( \theta ) &=  &e^{-i \pi / 4} \sum \frac{R_n e^{i n \theta}}{i -
R_n}
\end{eqnarray}

Weak scalar potentials  satisfy $R_n = R_{1-n}$. Then, for $\theta =
\pi$, we have $f ( \theta ) = 0$.
\subsection{Examples}
\subsubsection{Potential well} 
\begin{figure}[!t]
\begin{center}
\includegraphics[width=8cm,angle=0]{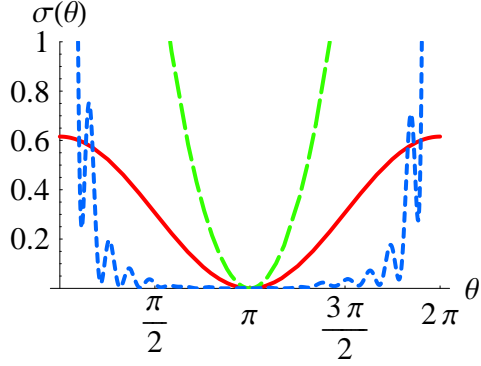}\\
\caption[fig]{(Color online). Dependence of the cross section on
  scattering angle for a circular potential well such that $k' r_0 = 1$. Red:
  $k r_0 = 0.1+k' r_0$. Long dashed (green): $k r_0 = 1+k' r_0$. Short dashed (blue): $k r_0 = 10+k' r_0$.}
\label{potential_fig}
\end{center}
\end{figure}

We assume that the potential for $r <
r_0$ is $V_0$. The wavevector of a state with energy $E = \vf k$ is
$k' = ( E + V_0 ) / \vf$. The matching conditions at $r = r_0$ are:
\begin{eqnarray}
J_n ( k r_0 ) + R_n Y_n ( k r_0 ) &= &T_n J_n ( k' r_0 ) \nonumber \\
J_{n+1} ( k r_0 ) + R_n Y_{n+1} ( k r_0 ) &= &T_n J_{n+1} ( k' r_0 )
\end{eqnarray}
and:
\begin{equation}
R_n = - \frac{J_n ( k r_0 ) J_{n+1} ( k' r_0 ) - J_{n+1} ( k r_0 )
  J_{n} ( k'
r_0 )}{Y_{n} ( k r_0 ) J_{n+1} ( k' r_0 ) - Y_{n+1} ( k r_0 ) J_{n}
(
  k' r_0 )}
\end{equation}
The largest values of $R_n$ when $r_0 \rightarrow 0$ are when $n=-1$
and $n=0$. We expand the Bessel functions, and obtain:
\begin{equation}
\sigma ( \theta ) \approx \frac{2 ( k' r_0 )^2}{\pi k \log^2 ( k r_0
)} \left[ 1 - \cos ( \theta ) \right]
\end{equation}
The cross section is zero for $\theta = \pi$, in agreement with the
suppression of backscattering in the absence of intervalley
transitions. Examples of the cross section due to a circular
potential are shown in Fig.[\ref{potential_fig}].
\subsubsection{Circular crack} 
\begin{figure}[!t]
\begin{center}
\includegraphics[width=8cm,angle=0]{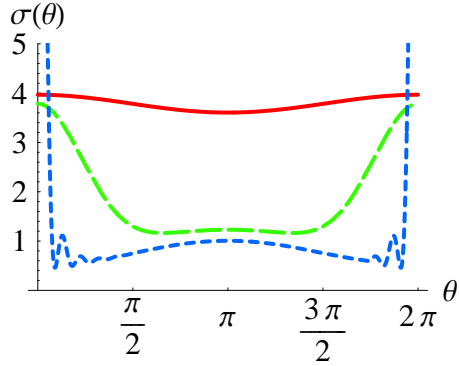}\\
\caption[fig]{(Color online). Dependence of the cross section on
  scattering angle for crack with zigzag edges. Red:
  $k r_0 = 0.1$. Long dashed (green): $k r_0 = 1$. Short dashed (blue): $k r_0 = 10$.}
\label{crack_fig}
\end{center}
\end{figure}

The boundary conditions for the
wavefunction $\Psi ( {\bf
  \vec{r}} ) \equiv [ \Psi_1 ( {\bf \vec{r}} ) , \Psi_2 ( {\bf \vec{r}} ) ]$ at
  void with zig-zag edges,
  satisfies $\Psi_1 ( {\bf \vec{r}} ) = 0$. Hence:
\begin{eqnarray}
J_n ( k r_0 ) + R_n Y_n ( k r_0 ) &= &0 \nonumber \\
R_n &= &- \frac{J_n ( k r_0 )}{Y_n ( k r_0 )}
\end{eqnarray}
The largest value of $R_n$ is for $n=0$. Unlike the previous case,
$R_0$ and $R_{-1}$ are not equal. To lowest order in $r_0$, the
cross section is:
\begin{equation}
\sigma ( \theta ) \approx \frac{\pi}{8 k \left[  \log ( k r_0 / 2)+
\gamma
  \right]^2}
\label{cross_crack}
\end{equation}
where $\gamma$ is the Euler-Mascheroni constant. The scattering
probability is isotropic. Examples of the angular dependence of the
cross section for a circular crack are shown in
Fig.[\ref{crack_fig}]. The behavior of the cross section in this
case is the same as that found for a strong scatterer in\cite{OGM06}

\subsubsection{Pentagonal cone.} 
\begin{figure}[!t]
\begin{center}
\includegraphics[width=8cm,angle=0]{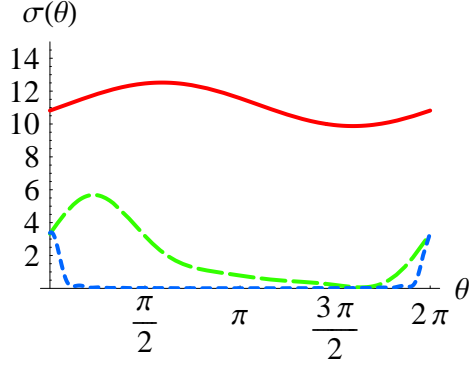}\\
\caption[fig]{\label{phaseshift}(Color online). Dependence of the
cross section on
  scattering angle for conical inclusion. Red:
  $k r_0 = 0.1$. Long dashed (green): $k r_0 = 1$. Short dashed (blue): $k r_0 = 10$.}
\label{cone_fig}
\end{center}
\end{figure}

When inserted into a graphene sheet,
a pentagon induces a disclination, and a point with finite
curvature. The sheet around a pentagon forms a cone. The cone can be
attached to a flat sheet with the inclusion of heptagons. We assume
that the boundary is sufficiently smooth, so that there is not
intervalley scattering. The flat surface is defined for $r \ge r_0$,
and the conical region for $r < r_0$. Inside the conical region, the
wedge induced by the dislocation is equivalent to a vortex at the
apex with flux $\Phi = \pi / 3$. The matching conditions are:
\begin{eqnarray}
J_{n} ( k r_0 ) + R_n Y_{n} ( k r_0 ) &= &T_n J_{n + \alpha} ( k r_0
) .\nonumber
\\
J_{n+1} ( k r_0 ) + R_n Y_{n+1} ( k r_0 ) &= &T_n J_{n + 1 +\alpha}
( k r_0 )
\end{eqnarray}
where $\alpha = - \Phi / ( 2 \pi ) = - 1 / 6$. Then:
\begin{equation}
R_n = - \frac{J_{n} ( k r_0 ) J_{n+1+\alpha} ( k r_0 ) -
J_{n+\alpha}
  ( k r_0
  ) J_{n+1} ( k
r_0 )}{Y_{n} ( k r_0 ) J_{n+1+\alpha} ( k r_0 ) - Y_{n+1} ( k r_0 )
  J_{n+\alpha} ( k r_0 )}
\end{equation}
The cases $n$ and $1-n$ are not equivalent, and the scattering cross
sections at angles $\theta$ and $2 \pi - \theta$ are not equal.
Examples are shown in Fig.[\ref{cone_fig}].

\subsection{Weak scatterers. The Born approximation}
\subsubsection{General framework}
 We expand the outgoing wave as:
\begin{equation}
\Psi_{out} ( {\bf \vec{r}} ) \approx \Psi_{in} ( {\bf \vec{r}} ) +
\int d^2 {\bf \vec{r}} G_0 ( {\bf \vec{r}} - {\bf \vec{r}}' , \omega
) {\cal V} ( {\bf
  \vec{r}}' ) \Psi_{in} ( {\bf \vec{r}}' )
\end{equation}
where both $G_0 ( {\bf \vec{r}} , \omega )$ and ${\cal V} ( {\bf
\vec{r}} )$ are $2 \times 2$ matrices, and $\omega$ is the energy of
the particle.

We also have:
\begin{equation}
lim_{r \rightarrow \infty} G_0 ( {\bf \vec{r}} , \omega ) =
\frac{1}{4 \pi^2} \int d^2 {\bf \vec{k}}
      \frac{e^{i {\bf
      \vec{k}} {\bf \vec{r}}} \left( \omega {\cal I} + \vf {\bf \vec{k}}
      \vec{\sigma} \right)}{\omega^2 - \vf^2 \left| {\bf \vec{k}}
      \right|^2} \sim
       \frac{\omega e^{i \omega r /
      \vf}}{\vf^2 \sqrt{2 \pi \omega r
      / \vf}}  \left(
      \begin{array}{cc} 1 & e^{i \theta} \\ e^{-i \theta}
      & 1 \end{array} \right)
\label{green}
\end{equation}
where $r = | {\bf \vec{r}} |$.
\begin{figure}[!t]
\begin{center}
\includegraphics[width=3cm,angle=0]{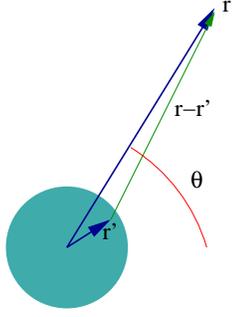}\\
\caption[fig]{\label{green_fig} (Color online). Notation used for
the calculation of the
  Green's
  function (see text for details).}
\label{green_geometry}
\end{center}
\end{figure}
We analyze the scattering from an angle defined by the unit vector
${\bf
  \vec{n}}'$ into an angle defined by ${\bf \vec{n}}$. Hence, the vector
  ${\vec{r}}$ is parallel to ${\bf \vec{n}}$. The labels defining the
  coordinates used for the calculation of the Green's function are sketched
  in Fig.[\ref{green_geometry}]. The incoming
  wavefunction is defined in eq.(\ref{wv_in}). At long distances, $| {\bf
  \vec{r}} | = r \rightarrow \infty$, we can expand:
\begin{equation}
\left| {\bf \vec{r}} - {\bf \vec{r}}' \right| \rightarrow r - {\bf
\vec{r}}' \bullet
    {\bf \vec{n}}
\end{equation}
where ${\bf \vec{n}} = {\bf \vec{r}} / r$. Defining $k_\omega =
\omega / \vf$, the Green's function becomes:
\begin{equation}
G ( {\bf \vec{r}} - {\bf \vec{r}'} , \omega ) \approx \frac{k_\omega
e^{i
    k_\omega (
    r  - {\bf \vec{n}} {\bf \vec{r}'} )}}{\vf \sqrt{2 \pi k_\omega r}} \left(
    \begin{array}{cc} 1 &e^{i \theta} \\ e^{-i \theta} &1  \end{array}
    \right)
\end{equation}
where $\theta$ is the angle approximately given  by the direction of
${\bf
  \vec{n}}$
\subsubsection{Scalar potential} We consider first a potential which
does not distinguish the two sublattices:
\begin{equation}
{\cal V} ( {\bf \vec{r}} ) \equiv \left( \begin{array}{cc} V ( {\bf
\vec{r}}
     ) &0
     \\ 0 &V ( {\bf \vec{r}} )  \end{array} \right)
\end{equation}
The outgoing wave is given by:
\begin{eqnarray}
\Psi_{out} ( r , \theta ) &\equiv &\int d^2 {\bf \vec{r}'}
\frac{k_\omega e^{i
    k_\omega (
    r  - {\bf \vec{n}} {\bf \vec{r}'} )}}{\vf \sqrt{2 \pi k_\omega r}} \left(
    \begin{array}{cc} V ( {\bf \vec{r}'} ) & V ( {\bf \vec{r}'}
    ) e^{i \theta} \\ V ( {\bf \vec{r}'} ) e^{-i \theta} & V (
    {\bf \vec{r}'} ) \end{array} \right) \times e^{ i k_\omega {\bf
    \vec{n}}_{in}
    {\bf \vec{r}'}} \left( \begin{array}{c} 1  \\ e^{ - i \theta_{in}}
    \end{array} \right) \nonumber \\ &= &\frac{k_\omega e^{i k_\omega
    r}}{\vf \sqrt{2 \pi k_\omega r}} \int d^2 {\bf \vec{r}'} e^{-i k_\omega (
    {\bf \vec{n}} - {\bf \vec{n}}_{in} ) {\bf \vec{r}'}} V ( {\bf
    \vec{r}'} )  \times \left( \begin{array}{c} 1 + e^{i ( \theta -
    \theta_{in} )} \\ e^{- i \theta} \left( 1 + e^{i ( \theta - \theta_{in}
    )} \right) \end{array} \right)
\end{eqnarray}
Finally, the amplitude which defines the scattering by an angle
$\theta - \theta_{in}$ is:
\begin{equation}
f ( \theta - \theta_{in} ) \propto \frac{\sqrt{k_\omega}}{\vf} \int
d^2 {\bf
    \vec{r}'} e^{- i k_\omega (
    {\bf \vec{n}} - {\bf \vec{n}}_{in} ) {\bf \vec{r}'}} V ( {\bf
    \vec{r}'}  ) \left[ 1 + e^{i ( \theta -\theta_{in}
    )} \right]
\end{equation}
and the cross section
\begin{equation}
\sigma ( \theta - \theta_{in} ) \propto \frac{k_\omega}{\vf^2}
\left| \int d^2 {\bf \vec{r}'} e^{- i k_\omega (
    {\bf \vec{n}} - {\bf \vec{n}}_{in} ) {\bf \vec{r}'}} V ( {\bf
    \vec{r}'} ) \right|^2
\left[ 1 + \cos ( \theta - \theta_{in} ) \right]
\end{equation}
and scattering in the backward direction, $\theta - \theta_{in} =
\pi$, is suppressed.

The previous calculation can be easily generalized to a potential:
\begin{equation}
{\cal V} ( {\bf \vec{r}} ) \equiv \left( \begin{array}{cc} \bar{V} (
{\bf
      \vec{r}} ) + \Delta V ( {\bf \vec{r}} ) &0 \\ 0 & \bar{V}  ( {\bf
      \vec{r}} ) - \Delta V ( {\bf \vec{r}} ) \end{array} \right)
\end{equation}
We find:
\begin{eqnarray}
f ( \theta - \theta_{in} ) &\propto &\frac{\sqrt{k_\omega}}{\vf}
\left\{ \int d^2 {\bf
    \vec{r}'} e^{-i k_\omega (
    {\bf \vec{n}} - {\bf \vec{n}}_{in} ) {\bf \vec{r}'}} \bar{V} ( {\bf
    \vec{r}'}  ) \left[ 1 + e^{i ( \theta -\theta_{in}
    )} \right] \right. \nonumber \\
&+ &\left. \int d^2 {\bf
    \vec{r}'} e^{i k_\omega (
    {\bf \vec{n}} - {\bf \vec{n}}_{in} ) {\bf \vec{r}'}} \Delta V ( {\bf
    \vec{r}'}  ) \left[ 1 - e^{i ( \theta -\theta_{in}
    )} \right] \right\}
\end{eqnarray}
The scattering is isotropic, independent of $\theta - \theta_{in}$,
when the potential modifies only one sublattice, $\bar{V} ( {\bf
\vec{r}} ) = \pm \Delta V ( {\bf \vec{r}} )$. Scattering in the
forward direction is suppressed when the potential is antisymmetric
in the two sublattices, $\bar{V} ( {\bf \vec{r}} ) = 0$.
\subsubsection{Isotropic elastic strains.}
An elastic distortion induces a gauge potential\cite{M07}:
\begin{equation}
{\cal V} ( {\bf \vec{r}} ) \equiv    \left( \begin{array}{cc} 0 &
\beta t
    \left( u_{xx} - u_{yy} + 2 i u_{xy} \right) \\ \beta t \left( u_{xx} -
    u_{yy} - 2 i u_{xy} \right) &0 \end{array} \right)
\end{equation}
where $u_{ij}$ are the components of the strain tensor, $t$ is the
nearest neighbor hopping term in the tight binding hamiltonian, and
$\beta = \partial t / \partial d$, where $d$ is the bond length.

An isotropic ripple with a height profile $h ( r )$ leads to (in
radial coordinales):
\begin{equation}
{\cal V} ( r , \phi ) \equiv    \left( \begin{array}{cc} 0 & \beta t
\left( \frac{\partial h}{\partial r} \right)^2 e^{- 2 i \phi}  \\
\beta t \left( \frac{\partial h}{\partial r} \right)^2 e^{ 2 i
\phi}&0 \end{array} \right)
\end{equation}
This potential is not isotropic, and the scattering does not depend
only on the angle  between the incoming and outgoing directions.  We
define
\begin{equation}
g ( r )= \beta t \left( \frac{\partial h}{\partial r} \right)^2
\end{equation}
The outgoing wave can be written as:
\begin{eqnarray}
\Psi_{out} ( r , \theta ) &\equiv &\int d^2 {\bf \vec{r}'}
\frac{k_\omega e^{i
    k_\omega (
    r  + {\bf \vec{n}} {\bf \vec{r}'} )}}{\vf \sqrt{2 \pi k_\omega r}} \left(
    \begin{array}{cc} g ( r ) e^{i ( \theta  + 2 \phi )} & g ( r ) e^{- 2 i
    \phi} \\
    g ( r ) e^{ 2 i \phi} &g ( r ) e^{i ( - \theta - 2 \phi)} \end{array}
    \right) \times \nonumber \\ &\times &e^{ i k_\omega {\bf
    \vec{n}}_{in}
    {\bf \vec{r}'}} \left( \begin{array}{c} 1  \\ e^{ - i \theta_{in}}
    \end{array} \right)
\label{wv_strain}
\end{eqnarray}
we can write:
\begin{equation}
( {\bf \vec{n}} - {\bf \vec{n}}_{in} ) {\bf \vec{r}'} = \left| {\bf
\vec{n}} -
  {\bf \vec{n}}_{in} \right| r' \cos ( \theta_{n-n_{in}} - \phi )
\end{equation}
where:
\begin{equation}
\tan ( \theta_{n-n_{in}} ) = \frac{\sin ( \theta ) - \sin (
\theta_{in}
  )}{\cos ( \theta ) - \cos ( \theta_{in} )} = - \cot \left( \frac{\theta +
    \theta_{in}}{2} \right)
\end{equation}
so that:
\begin{equation}
 \theta_{n-n_{in}} = \frac{\theta + \theta_{in}}{2} - \frac{\pi}{2}
\end{equation}
we can make the change of variables $\phi = \phi' + ( \theta +
\theta_{in} ) / 2 - \pi / 2$ in the integral over $\phi$ in
eq.(\ref{wv_strain}), and obtain:
\begin{eqnarray}
\Psi_{out} ( r , \theta ) &\equiv &\int r' dr' d \phi'
\frac{k_\omega e^{i
    k_\omega (
    r  )}}{\vf \sqrt{2 \pi k_\omega r}} e^{i k_\omega | {\bf \vec{n}} - {\bf
    \vec{n}'}
    | r' \cos{\phi'}} g ( r'
    ) \times \nonumber \\ &\times &\left(
    \begin{array}{c}  e^{i ( 2 \phi' + 2 \theta + \theta_{in}+ \pi )} +
    e^{i (- 2 \phi' - \theta  - 2 \theta_{in} - \pi )} \\
     e^{i ( 2 \phi' +  \theta +\theta_{in} + \pi )} + e^{i ( - 2
     \phi' - 2 \theta - 2 \theta_{in} - \pi )} \end{array}
    \right) \nonumber \\
&= & \int d r'  \frac{k_\omega e^{i
    k_\omega (
    r  )}}{\vf \sqrt{2 \pi k_\omega r}} J_2 ( k_\omega | {\bf \vec{n}} - {\bf
    \vec{n}'}  | r' ) g ( r' ) \times \nonumber \\ &\times &\left( \begin{array}{c} e^{- i \theta_{in}}
     \left( e^{2 i ( \theta + \theta_{in} )} + e^{- i
     ( \theta + \theta_{in} ) } \right) \\ e^{- i \theta} e^{- i \theta_{in}}
    \left( e^{2 i ( \theta + \theta_{in} )} + e^{- i
     ( \theta + \theta_{in} ) } \right)
\end{array} \right)
\end{eqnarray}
where $J_2 ( x )$ is a Bessel function. The scattering amplitude can
be written as:
\begin{equation}
f ( \theta , \theta_{in} ) \propto -  \frac{\sqrt{k_\omega}}{\vf}
e^{- i
     \theta_{in}}
     \left( e^{2 i ( \theta + \theta_{in} )} + e^{- i
     ( \theta + \theta_{in} ) } \right) \int d r'  J_2 ( k_\omega | {\bf
     \vec{n}} - {\bf
    \vec{n}'}  | r' ) g ( r' )
\end{equation}
and the cross section is:
\begin{equation}
\sigma ( \theta , \theta_{in} ) \propto \frac{k_\omega}{\vf^2}
\left|  \int d
    r'  J_2 ( k_\omega | {\bf \vec{n}} - {\bf
    \vec{n}'}  | r' ) g ( r' ) \right|^2 \left\{ 1 + \cos [ 3 ( \theta +
    \theta_{in} ) ] \right\}
\label{cross_ripple}
\end{equation}
This function reflects the threefold symmetry of the honeycomb
lattice.

Near the Dirac point, $| k_\omega | \rightarrow 0$, the cross
section in eq.(\ref{cross_ripple}) grows as $| k_\omega |^3$. For a
ripple of height $h$ and size $l$, we have:
\begin{equation}
g ( r ) \approx \beta t \times \left\{ \begin{array}{cc} \left(
\frac{h}{l}
    \right)^2 &r \ll l \\ 0 & r \gg l \end{array} \right.
\end{equation}
The hopping $t$ is given approximately, by $t \approx \vf / a$,
where $a$ is the interatomic distance. The total cross section,
obtained by integrating eq.(\ref{cross_ripple}) over angles is given
by:
\begin{equation}
\sigma \sim \left\{ \begin{array}{cc} \frac{\beta^2 h^4 k_\omega (
    k_\omega l )^4}{a^2}  & k_\omega l \ll 1 \\ \frac{\beta^2 h^4
    k_\omega}{a^2} &
    k_\omega l \sim 1 \\ \frac{\beta^2 h^4 k_\omega}{a^2 ( k_\omega l )^4} &
    k_\omega l \gg 1 \end{array} \right.
\end{equation}
The total cross section has a maximum for $k_\omega \sim l^{-1}$,
$\sigma_{max} \sim ( \beta t /\vf )^2 h^4 / l$. For $h \sim 1$nm and
$l \sim 10$nm, we obtain $\sigma_{max} \sim 10 h \sim 10$nm.
\subsubsection{Effective magnetic vortex}
Topological lattice defects, such as disclinations and dislocations
induce, in addition to long range strains, an effective vortex at
their core\cite{GGV92,GGV93b,MG06}, which mixes the two valleys in
the case of a disclination. We consider now the scattering by such a
vortex alone, neglecting the effect of the elastic strains. An
isotropic distribution of a (fictitious) magnetic field induces a
potential:
\begin{equation}
{\cal V} ( r , \phi ) \equiv \left( \begin{array}{cc} 0 & g ( r )
e^{i \phi}
    \\ g ( r ) e^{- i \phi} & 0 \end{array} \right)
\end{equation}
where $g ( r ) \propto ( e / c ) \vf r B ( r )$. Using the same
scheme as in the previous case, we find:
\begin{eqnarray}
\Psi_{out} ( r , \theta ) &\equiv &\int d^2 {\bf \vec{r}'}
\frac{k_\omega e^{i
    k_\omega (
    r  + {\bf \vec{n}} {\bf \vec{r}'} )}}{\vf \sqrt{2 \pi k_\omega r}} \left(
    \begin{array}{cc} g ( r ) e^{i ( \theta  - \phi )} & g ( r ) e^{ i
    \phi} \\
    g ( r ) e^{ - i \phi} &g ( r ) e^{i ( - \theta + \phi)} \end{array}
    \right) \times \nonumber \\ &\times &e^{ i k_\omega {\bf
    \vec{n}}_{in}
    {\bf \vec{r}'}} \left( \begin{array}{c} 1  \\ e^{ - i \theta_{in}}
    \end{array} \right) = \nonumber \\
&= & i \int d r'  \frac{k_\omega e^{i
    k_\omega (
    r  )}}{\vf \sqrt{2 \pi k_\omega r}} J_1 ( k_\omega | {\bf \vec{n}} - {\bf
    \vec{n}'}  | r' ) f ( r' ) \times \nonumber \\ &\times &\left( \begin{array}{c} e^{i ( \theta -
    \theta_{in} + \pi ) / 2} + e^{i ( \theta - \theta_{in}  + \pi ) / 2} \\
    e^{- i ( \theta + \theta_{in} + \pi ) / 2} + e^{-i ( \theta + \theta_{in}
    - \pi ) / 2}
\end{array} \right)
\label{wv_field}
\end{eqnarray}
so that:
\begin{equation}
f (  \theta - \theta_{in} ) \propto e^{i ( \theta - \theta_{in} ) /
2} \frac{\sqrt{k_\omega}}{\vf} \int d r' J_1 ( k_\omega | {\bf
\vec{n}} - {\bf
    \vec{n}'}  | r' ) g ( r' )
\end{equation}
We assume that:
\begin{equation}
g ( r ) \approx \left\{ \begin{array}{cr} f \times \frac{\vf
\Phi_0}{l} &r
    \ll l \\
0 &r \gg l \end{array} \right.
\end{equation}
where $f$ is a dimensionless number of order unity, $\Phi_0$ is the
flux quantum, and $l$ is the redius of the distorted region, we find
that the total cross section behaves as:
\begin{equation}
\sigma ( k_\omega ) \sim \left\{ \begin{array}{cl} k_\omega f^2 l^2
(
    k_\omega l )^3 &k_\omega l \ll 1 \\  k_\omega f^2 l^2 & k_\omega l \sim 1
    \\ \frac{ k_\omega f^2 l^2}{( k_\omega l )^3}  &
    k_\omega l \gg 1 \end{array} \right.
\end{equation}
\subsection{Lattice effects.}
\subsubsection{Green's function formulation}
\begin{figure}[!t]
\begin{center}
\includegraphics[width=4cm,angle=0]{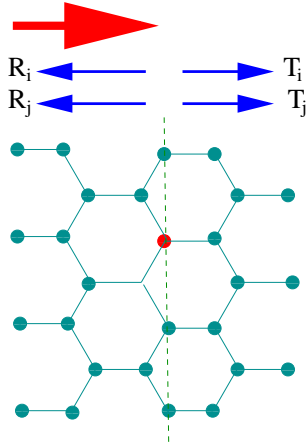}\\
\caption[fig]{(Color online). Geometry used to calculate scattering
  amplitudes
  in the discrete honeycomb lattice. See text for details.}
\label{lattice_fig}
\end{center}
\end{figure}

\begin{figure}[!t]
\begin{center}
\includegraphics[width=8cm,angle=0]{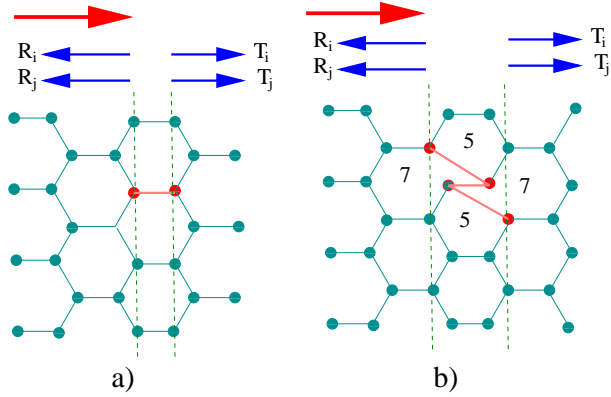}\\
\caption[fig]{(Color online). Geometry used to calculate scattering
  amplitudes
  in the discrete honeycomb lattice in the presence of bond disorder. Left:
  perturbation of a single bond. Right: Stone-Wales defect. See
  text for details.}
\label{lattice_bond_fig}
\end{center}
\end{figure}

We can generalize the previous analysis to the discrete honeycomb
lattice. The scattering is no longer isotropic, and only some
incident angles can be studied analytically. We fix the direction of
the incident wave, and use periodic boundary conditions in the
perpendicular direction. A particular case, where the incident wave
is along one of the symmetry axes of the honeycomb lattice is
sketched in Fig.[\ref{lattice_fig}]. The analysis requires the
calculation of the transmission and reflection coefficients for the
different channels in the problem. The scheme can easily be
generalized to any local lattice defect, as sketched in
Fig.[\ref{lattice_bond_fig}]. This scheme, using the continuum
approximation for the local Green's function, has been discussed
in\cite{OGM06}.
The scattering process can be solved if the amplitudes of the
reflected and transmitted waves along each of the transverse
channels of the lattice can be determined. We label the transverse
channels by the angular momentum $k_\perp$. Using periodic boundary
conditions with $N$ unit cells in the transverse direction, the
allowed values of $k_\perp$ are $k_\perp = 2 \pi n / ( N a ) , n = 0
, \cdots , N-1$; $a$ is the lattice constant.

We analyze next the scattering by a local impurity, and then extend
the calculation to more complex lattice defects.
\subsubsection{Scattering by an impurity.} A local impurity perturbs
only one site of the lattice. The matching of the incoming and the
reflected and transverse waves needs only be defined along a line of
sites which includes the vacancy. We assume that the incoming wave
has $k_\perp^0 = 0$. There are $N$ reflection and transmission
amplitudes, $R_{k_\perp} , T_{k_\perp}$. The continuity of the
wavefunctions allows us to define $N$ equations:
\begin{eqnarray}
1 + R_0 &= &T_0 \nonumber \\
R_{k_\perp} &= &T_{k_\perp} \, \, \, \, \, \, \, \, \, \, k_\perp
\ne 0 \label{trans_1}
\end{eqnarray}
The other $N$ equations needed to determine uniquely the values of
$R_{k_\perp} , T_{k_\perp}$ are given by the condition that the full
wavefunction must be an eigenvector with energy $\epsilon$: We use
eq.(\ref{trans_1}) in order to eliminate the $R_{k_\perp}$, and
obtain:
\begin{eqnarray}
G_0^{-1} ( k_\perp , 0 , \epsilon ) \times T_{k_\perp} & = &G_0^{-1}
( k_\perp , 0 , \epsilon ) + V \sum_{k_\perp'} G_0 ( k_\perp' , 0 ,
\epsilon ) \, \, \, \, \, \, \, \, \, \, k_\perp =
0 \nonumber \\
G_0^{-1} ( k_\perp , 0 , \epsilon ) \times T_{k_\perp} & = & V
\sum_{k_\perp'} G_0 ( k_\perp' , 0 , \epsilon ) \, \, \, \, \, \, \,
\, \, \, \, \, \, \, \, \, \, \, \, \,k_\perp \ne 0
\label{transmission}
\end{eqnarray}
where $V$ is the strength of the impurity potential, and:
\begin{equation}
G_0 ( k_\perp , 0 , \omega  ) = \sum_{k_\parallel} \frac{1}{\omega -
\epsilon
  ( k_\parallel , k_\perp )}
\label{green_row}
\end{equation}
is the Green's function of the problem, resolved in transverse
momentum, and projected on the row where the impurity is located.
The energies $\epsilon ( k_\parallel , k_\perp )$ correspond to the
unperturbed hamiltonian.

The solution of eq.(\ref{transmission}) is:
\begin{eqnarray}
T_{k_\perp} & = &1 + \frac{V G_0 ( k_\perp , 0 , \epsilon )}{1 - V
  \sum_{k_\perp'} G_0 ( k_\perp' , 0 , \epsilon ) }\, \, \, \, \, \, \, \, \,
  \,
\, \, \,
\, \, \, \, \, \, \,k_\perp = 0 \nonumber \\
T_{k_\perp} & = &\frac{V G_0 ( k_\perp , 0 , \epsilon )}{1 - V
  \sum_{k_\perp'} G_0 ( k_\perp' , 0 , \epsilon ) }\, \, \, \, \, \, \, \, \,
  \, \, \, \, \, \, \, \,
\, \, \, \, \, \, \,k_\perp \ne 0 \label{transmission_lattice}
\end{eqnarray}
At low energies, we can write:
\begin{equation}
\epsilon ( k_\parallel , k_\perp ) \approx \vf \sqrt{ k_\parallel^2
+
  k_\perp^2 }
\end{equation}
and:
\begin{eqnarray}
G_0 ( k_\perp , 0 , \omega ) &\approx &\frac{k_\parallel}{\vf \sqrt{
  k_\parallel^2 + k_\perp^2 }} \approx \frac{1}{\vf \cos ( \theta )}
  \nonumber \\
\sum_{k_\perp'} G_0 ( k_\perp' , 0 , \epsilon ) &\approx
&\frac{\sqrt{
  k_\parallel^2 + k_\perp^2}}{\vf} \left[ i + \log \left(
  \frac{\Lambda}{\sqrt{ k_\parallel^2 + k_\perp^2 }} \right) \right]
\end{eqnarray}
where $\Lambda$ is a high momentum cutoff of the order of the
inverse of the lattice constant, $\Lambda \sim a^{-1}$.

In order to obtain the total transmitted current as function of the
incoming current one must multiply $| T_{k_\parallel} |^2$ by the
outgoing current, $j_{k_\parallel}$ and divide it by the ingoing
current, $j_0$. In the tight binding model which describes each
transverse channel, we have that $j_\parallel = G_0^{-1} ( k_\perp ,
0 , \omega )$. Hence, we can write for the cross section:
\begin{equation}
\sigma ( k_\parallel , \epsilon ) \propto \frac{G_0 ( k_0 , \epsilon
) V^2
  G_0 ( k_\parallel , \epsilon )}{| 1 - V  \sum_{k_\perp'} G_0 ( k_\perp' , 0
  , \epsilon ) |^2}
\end{equation}
and, in order to obtain the dependence of the cross section on the
outgoing angle, $\theta$, we must take into account that:
\begin{equation}
\sigma ( \theta , \epsilon ) d \theta = \sigma ( k_\parallel ) d
k_\parallel = \sigma ( k_\parallel , \epsilon ) \sqrt{k_\parallel^2
+ k_\perp^2} \cos ( \theta ) d \theta
\end{equation}

Using these expressions, we find that the scattering by a lattice
impurity is isotropic at low energies, and valley independent (note
that the calculation does not distinguish the valley index of the
transmitted wave) . For $V \ll t$, where $t$ is the nearest neighbor
hopping, the total cross section is given, approximately, by:
\begin{equation}
\sigma \sim \frac{\sqrt{k_\parallel^2 + k_\perp^2} V^2 a^4}{\vf^2}
\end{equation}
and, for $V \gg t$, we find, neglecting logarithmic corrections:
\begin{equation}
\sigma \sim \frac{1}{\sqrt{k_\parallel^2 + k_\perp^2}}
\end{equation}
In agreement with the results obtained for a crack in the continuum
limit, eq.(\ref{cross_crack}).
\section{Conclusions}
We have analyzed scattering processes which will affect the mobility
of carriers in graphene. We show that localized defects can be
classified into at least two types with opposite dependence of the
cross section on density: i) weak scatterers, where the cross
section grows as the square root of the density, and strong
scatterers, where the cross section decreases as the square root of
the density. The resulting conductivity can be either density
independent, or grow linearly with density.

We have also studied defects with internal structure, such as those
induced by elastic strains or ripples, where the perturbation
couples to the Dirac quasiparticles as an effective gauge field. We
find that for ripples with a characteristic size $l$, the cross
section is highest for a density such that $\kf^{-1} \sim l$. The
scattering at each valley does not show a symmetry between $\theta$
and $- \theta$, where $\theta$ is the incident angle, as expected
from general symmetry considerations.

We have finally shown that the classification of short range
potentials into weak and strong scatterers with different dependence
on carrier density remains unchanged, even when intervalley
scattering is important.
\section{Acknowledgements}
This work was supported by MEC (Spain) through grant
FIS2005-05478-C02-01, the Comunidad de Madrid, through the program
CITECNOMIK, CM2006-S-0505-ESP-0337, and the European Union Contract
12881 (NEST).

\bibliography{bib_scattering}

%
%

\end{document}